\documentstyle[11pt,aaspp4]{article}

\def\etal{{\it et al.}}
\def\ie{{\it i.e.}}

\lefthead{Ledlow and Owen} 
\righthead{Radio/Optical Luminosity Functions}

\begin{document}

\title{A 20 cm VLA SURVEY OF ABELL CLUSTERS OF GALAXIES\\
    VI. RADIO/OPTICAL LUMINOSITY FUNCTIONS}

\author{Michael J. Ledlow\altaffilmark{1,2}}
\affil{New Mexico State University, Dept. of Astronomy,\\
   Las Cruces, New Mexico  88003}
\and

\author{Frazer N. Owen}
\affil{National Radio Astronomy Observatory\altaffilmark{3}, Socorro, New Mexico 87801}

\altaffiltext{1}{Also National Radio Astronomy Observatory, Socorro, New Mexico}
\altaffiltext{2}{Also University of New Mexico, Institute for Astrophysics,
Albuquerque, New Mexico}
\altaffiltext{3}{The National Radio Astronomy Observatory is operated by 
Associated Universities, Inc., under cooperative agreement with the National
Science Foundation} 

\begin{abstract}

From a statistically complete sample of 188 radio galaxies in Abell clusters,
we examine the radio/optical correlations, the FR I/II division, and the
univariate and bivariate luminosity functions.  As suggested by Owen (1993), 
the FR I/II division is shown
to be a strong function of the optical luminosity of the host galaxy 
($\propto L_{opt}^2$).  This dependence is also seen in the bivariate 
luminosity function, which suggests that the evolutionary tracks
of radio sources and/or the initial conditions in the source are governed
by the host galaxy properties.  The probabilty for detecting
radio emission increases with optical luminosity.  The optical dependence
is clearly separated in the integral luminosity functions which can be 
used as a constraint to models of FR I radio power evolution.  Additionally, 
the source counts from the integrated univariate radio luminosity function 
(RLF) are consistent with our suggestion in paper V that radio sources may be 
a transient phenomenon which occurs in all elliptical galaxies at some 
time (or several times) over their lifetime.  We find no statistically 
significant differences in the luminosity functions between rich cluster
samples and radio sources not selected to reside in clusters.  These 
results suggest that all radio galaxies live in similar environments
in that the optical luminosity and the properties of the host galaxy 
are the most important parameters which affect radio source formation
and evolution. 

\end{abstract}

%\keywords{clusters: globular, peanut --- bosons: bozos}

% In the first two sections, you should notice the use of the LaTeX \cite
% command to identify citations.  The citations are tied to the
% reference list via symbolic KEYs.  We have chosen the first three
% characters of the first author's name plus the last two numeral of the
% year of publication.  The corresponding reference has a \bibitem
% command in the reference list below.
%
% Please see the AASTeX manual for a more complete discussion on how to make
% \cite-\bibitem work for you.   

\section{Introduction}

     This is the sixth paper devoted to the analysis of a VLA 20cm 
survey of over 500 Abell clusters of galaxies.  In paper I-III 
(Zhao \etal\ 1989, Owen \etal\ 1992, Owen \etal\ 1993), the VLA C-Array
survey was presented, including optical ID's, radio maps, and source 
parameters.  In paper IV (Ledlow and Owen, 1995a), the final statistical
sample was assembled along with an analysis of the cluster properties
and their effect on the radio source detection rates.  In paper V 
(Ledlow and Owen, 1995b), R-Band CCD surface photometry for 265 radio
galaxies was presented, and compared to the optical properties of a 
control sample of radio-quiet galaxies chosen from the same clusters. 
In this paper, we examine the relationship between radio and optical 
luminosity and compute the univarite radio and bivariate radio/optical
luminosity functions for our statistical sample of 188 radio galaxies 
with $z<0.09$.  These functions are compared to other samples of objects
not selected from rich-clusters to examine the effects of the environment
on radio source populations.  

    Owen and Laing (1989) examined the optical properties of 47 $3CR$ or 
other non-rich cluster radio galaxies.  Owen and White (1991) followed 
up this study with optical observations of 49 rich cluster radio galaxies.
While no significant differences were found between the cluster/non-cluster
objects, using the larger, combined data set, Owen and White suggested
that the FR I/II sources were not only separated by a radio power division, 
but that this {\it break} appeared to depend on the optical luminosity 
of the host galaxy.  Owen (1993) followed up this idea by combining
optical magnitudes from several samples from the literature, including 
objects from a variety of environments and over a large-range in redshift
($z<0.5$).  Owen found that the separation between the FR I/II classes 
was very sharp and was strongly dependent on the optical luminosity ($\propto 
L_{opt}^2$).  He suggested that the sharpness of this break may indicate
some relationship between the FR classes, possibly an evolutionary relation.  
Owen and Ledlow (1994) presented preliminary determinations of the 
bivariate luminosity functions from the statistically complete VLA sample 
(Ledlow and Owen, 1994a). They showed that the break in the luminosity functions 
corresponded exactly to the FR I/II division
and had the same optical dependence.  It is therefore apparent that the 
properties of the host galaxy must obviously influence the radio morphology.   

    In Paper V, using optical surface photometry of 265 cluster radio galaxies and
50 radio-quiet ellipticals from the same clusters, we determined that there
is no significant difference in the optical properties between radio-loud (FR I) and
quiet elliptical galaxies.  This result is also consistent with Owen, Ledlow, and Keel
(1995) from analysis of the optical spectra and comparison to the radio-quiet
sample of Phillips \etal\ (1986).  The optical luminosity is the fundamental 
parameter which describes elliptical galaxies as a class.  All other optical 
measures (slope of the surface-brightness profile, ellipticity, size, etc.)
can be factored by the total optical luminosity.  These results suggest that
there is not a unique population of ellipticals which form radio sources.  
However, from the optical dependence of the FR I/II break, the host galaxy 
properties must influence the morphology and observed properties of 
the radio source. 

    Several papers have addressed the univariate radio luminosity function for 
non-cluster objects (Cameron, 1971; Colla \etal\ 1975; Parma \etal\ 1986;
de Ruiter \etal\ 1986; Fanti \etal\ 1986,1987). 
Most of these studies have used radio surveys
such as the Bologna B2 survey or the Cambridge 3CR sample of powerful
radio galaxies.  These samples have been compared to observations of 
rich cluster radio galaxies by Jaffe and Perola (1976), Auriemma \etal\ (1977),
Lari and Perola (1978), Fanti \etal\ (1982), and Fanti (1984).  These
studies found no difference in the univariate radio luminosity functions
for those samples selected in and out of rich clusters (although the sample
size of rich clusters was small).  The conclusions
are that the large-scale environment does not change the probability
of detecting radio sources and does not affect radio source lifetimes.
This has been a puzzling discovery given that the 
density of galaxies in and out of rich clusters can differ by an order 
of magnitude.  Additionally, gas densities inferred from X-ray observations
of clusters (Sarazin, 1986), infer possible differences of $>100$ between poor clusters
or the field and rich clusters.  Thus, the aspects of the environment which
which are important to the formation and evolution of radio sources 
is not well understood.

     Very few studies have had sufficient data to calculate the bivariate
radio/optical luminosity functions.  Existing data has been limited to 
Zwicky magnitudes, which suffer fairly large systematic errors $\geq 0.3$
magnitudes (Colla \etal\ 1975).  Auriemma \etal\ (1977) were the first
to suggest that the probabilty for radio emission increased with optical 
luminosity.  However, with such large errors in the optical magnitudes, 
the exact dependence on the optical luminosity was difficult to determine. 
Sadler \etal\ (1989) also calculated bivariate luminosity functions 
for a VLA survey of non-cluster radio galaxies.  In this paper, we will
compare our results to these two studies to examine how the optical luminosity 
and the environment affect radio source detection and properties.  
Preliminary results from the bivariate luminosity functions for our 
cluster sample were reported in Owen and Ledlow (1994).  

    In section 2, we discuss the relationship between radio and optical 
luminosity. 
In section 3, we outline the procedure for calculating and normalizing 
the luminosity functions.  In section 4, we present the 
differential and integral univariate and bivariate luminosity functions
for our cluster sample.  We compare the luminosity functions from 
our cluster sample to other non-cluster samples and discuss the importance
of the environment in section 5.  We summarize our conclusions in section 6.

     Throughout this paper we assume $H_0 = 75~\rm km~sec^{-1}~Mpc^{-1}$ 
and $q_0 = 0$. 

\section{Radio/Optical Properties}

     Radio galaxies can be coarsely grouped
into two classifications : {\sl FR I} and {\sl FR II's}.  These
classifications were defined by Fanaroff and Riley \markcite{fana1974}
(1974).  See Bridle (1984), Heckman \etal\ (1994), and Baum \etal\ (1995) for 
a more detailed description of FR I and FR II radio galaxies.  
Simply stated,
the {\sl FR I's} are dominated by emission from the compact core and jets.
The outer regions of the source consist of diffuse
lobes which fade with distance. The {\sl FR II's} have the highest
radio brightness far from the host galaxy.  These hotspots are thought to
be coincident with the location of the working surface of the beam, 
and the radio lobes are
the swept-back material or backflow from the shocked region in the wake 
of the advance of the head of the jet.
The one observed difference between these two classifications is the total
radio power.  Fanaroff and Riley noticed that the {\sl FR II's} occurred
almost exclusively at radio powers $> 10^{24-25}~W~Hz^{-1}$.  The
{\sl FR I's} almost always have radio powers below this value.  {\sl FR I's}
are usually interpreted as subsonic or transonic entraining flows (on kpc scales).
The {\sl FR II's} are thought to have supersonic jets.  However, VLBI observations
suggest that both FR I's and FR II's have similar relativistic flows on parsec 
scales ({\it e.g.\/} Pearson, 1996)

     In this paper, we will divide our sample of radio
galaxies into only these two divisions.
Owen and Laing \markcite{owen1989}(1989) and
Owen and White \markcite{owen1991}(1991) further divide the {\sl FR I} class into Twin-Jets (TJ),
Wide-Angle Tails (WAT), Narrow-Angle Tails (NAT), and Fat-Doubles (FD).
There are, of course, many sources which are difficult to classify uniquely.  For 
purposes of this paper, we identify sources as FR II only if they meet the 
strict definition of Fanaroff and Riley (1974).  All other sources are put in 
the FR I class. 

     In paper V (Ledlow and Owen, 1995b), we found that cluster radio 
galaxies are not
obviously different from normal cluster ellipticals.  However, once a
galaxy forms a radio source, how do the properties of the host galaxy
affect the observed radio properties?  The optical properties 
are each a measure of different physical parameters within the host galaxy.
The optical luminosity is a measure of
the mass of the system and thus may be related to the efficiency or
fueling of the active nucleus.  In addition, higher mass objects will
have a deeper local gravitational potential.  The interstellar gas
density and pressure as well as the extent of the ISM, and possibly the
mass and size of the gaseous halo surrounding the galaxy are also a 
function of the optical luminosity.  
We therefore might expect that the local dynamics which govern the conditions
in the environment in which the radio jets propogate would influence the
morphology or possibly the radio power output of the source.   

     The statistics from our VLA cluster survey are that 94\% of the radio
galaxies have FR I morphology.    In the lower redshift group ($z<0.09$) 
there are
only 3 {\sl FR II's} in the sample.  {\sl FR II's} are found 
in rich clusters at higher redshifts however.  They are also found
in bright galaxies not located in rich clusters.
Prestage and Peacock (1988) calculated the angular cross-correlation 
function for the two types, and found that 
{\sl FR I's} are found in much higher density groups.  FR II's seem to 
prefer approximately a factor of four lower galaxy density.  
One interpretation is that galaxies and the gaseous environment in rich clusters was
different in the past. 
Alternatively, independent of the clustering environment, 
the scarcity of FR II's may be a volume-effect.  
The most powerful
radio galaxies are, by definition, the rarest objects because of the steep
slope of the radio luminosity function.  When the search volume is increased
(by extending the surveys to higher redshifts), more of the powerful sources
are found.   It is unclear which effect (evolution in the cluster environment, 
galaxy density, or the search volume) is the dominant factor in FR II number 
counts. 

     The argument based on evolution with epoch is difficult to address from
our low-redshift sample.  Our sample was purposely chosen to examine a population
of radio galaxies unaffected by cosmological evolution, but rather to observe
a large number of radio galaxies in different stages of their individual
evolution.  More observations of high redshift cluster radio galaxies is
necessary to sort out these effects.

     We will first examine the radio/optical luminosity properties of the
sample.  This importance of this diagram was first discussed by Owen (1993).   
We have reproduced the radio/optical luminosity diagram from Owen 
\markcite{owen1993}(1993)
which combines the samples from Owen and Laing (1989), Owen and White (1991),
and several others from the 
literature.  This diagram is shown in figure 1.  The {\sl 1's} and
{\sl 2's} refer to the {\sl FR} classifications.  The other data
sets included in the diagram are from Lilly and Prestage \markcite{lill1987}
(1987), Smith
and Heckman \markcite{smit1989}(1989),
Baum \etal\ \markcite{baum1988}(1988), Hill and Lilly \markcite{hill1991}
(1991), and a portion
of the data from our current radio galaxy sample (Owen \etal\ \markcite{owen1992}1992).
These data are not based on a complete sample, but rather were chosen
to populate all regions in the diagram.  These data includes objects  
both in and out of rich clusters as well as spanning a range in redshifts
from $0.01$ to $0.5$.  All objects fit the observed relation nicely
despite their very different selection.  We see very clearly that the
two {\sl FR} classes populate this plane in different ways.  The {\sl FR}
division also appears to have a non-zero slope in the diagram.  An
approximate interpolation between the two classes gives a slope of $-0.66$
which corresponds to $L_{radio} \propto L_{opt}^{1.8}$.  The {\sl FR I/II}
division is quite sharp but appears to occur over a range in radio power
from $10^{24-26.5}~W~Hz^{-1}$ over the plotted range in optical luminosity.
This is very interesting in that many {\sl FR I's} are seen at radio
powers equivalent to {\sl FR II} sources, but in galaxies one to two
magnitudes brighter.  This fact coupled with the sharpness of the
division makes it seem unlikely that the two classes are totally unrelated
without some connection (possibly evolutionary?).  One would not expect
two uncorrelated samples to exhibit a sharp division in their optical/radio 
properties.  The optical dependence of this division is puzzling. 
It is not clear how the jet-producing AGN at the center of the galaxy 
can know about the total luminosity or mass.  If the initial conditions in 
the jets are similar in FR I's and FR II's, however, 
the optical dependence may be related to environmental factors of the host 
galaxy ISM (Bicknell 1995).  This would suggest a fundamental difference 
in the host galaxy properties at the same optical luminosity (see Baum, Zirbel, 
\& O'Dea 1995 for related discussions on this topic). 

    In figure 2 is plotted an identical diagram for our
entire cluster sample.  This plot includes the higher redshift sample
($0.09<z<0.25$) (see Ledlow and Owen, 1995a). 
Also plotted is the approximate fit to the {\sl FR}
division from figure 1.  We see that the {\sl FR} division
is much less clear in this plot.  There is a fair amount of overlap
across this line.  All the {\sl FR II's} which are below the division 
may be somewhat unusual.  Whereas most {\sl FR II's} are large, and extend well 
into the intracluster/intragroup medium, 
nearly all of these sources appear to be confined within the optical extent 
of the host-galaxy.  These objects may be
a class of {\sl FR II's} which have been missed in flux-limited surveys
of high-power radio galaxies (such as the {\sl 3CR} sample).  Their lower powers
and small size may indicate that they are young objects.  If {\sl FR II's}
are powered by supersonic flows, the amount of time spent in this state
would be small.  Therefore these objects may have been observed in a
rather rare stage of their evolution.  Alternatively, because all of these
unusual {\sl FR II's} are located in bright galaxies at the centers of 
rich clusters, the higher density ISM/ICM may be more efficient at confining the
radio source and limiting the size and total radio power. 

     The lack of any
high-power {\sl FR II's} in the upper right portion of the diagram may
be a selection effect.  Because the radio luminosity function falls rapidly
for powers above $10^{25}$ (see figure 6 and \S\ 4),
the number of objects in this radio power
range is very small.  Also, the optical luminosity function is very steep
in this range (see figure 4).  Thus at low
redshift one would expect few objects in these optical/radio luminosity
ranges because of the small search volume.  At much higher redshifts
($\sim 0.5$), the larger volume ($\propto$ distance$^{3}$) increases 
the probability of detecting these objects.  
It has often been cited in the literature that {\sl FR II} sources
are not found in rich cluster environments.  While clearly
they are a minority,
they are not totally absent from the centers of rich clusters.  The large
majority of the objects are however {\sl FR I's}.

     In figure 3, we plot the same diagram restricted to
our statistically complete sample for $z<0.09$. We see the same trend
seen in the previous diagrams, but notice that the {\sl FR II's} are
almost completely absent from this sample.
If we examine the distribution in radio/optical
luminosities from this figure, we see that for any given optical luminosity
there is a large range in observed radio power.  Depending on
where the radio flux-limit is chosen, different results might be observed.
For example, if one were to look only at the sources with powers $> 10^{23.5}$,
the distribution of the {\sl FR I's} in this plane would appear to
correlate with optical power $\sim L_{opt}^{2}$.  As we see from the fit
to the {\sl FR I/II} division, this appears to simply be an artifact
from selecting galaxies near the {\sl FR I/II} break.  When we include the
lower radio power sources, the previously strong $L_{opt}/L_{radio}$
correlation disappears.  
So while it is suspected that
the probability for a galaxy to be a radio source is an increasing
function of optical luminosity, there is no strong correlation
between individual radio and optical luminosities.  However, because we are 
observing sources in all stages of their evolution, one might not expect 
to see a strong correlation.  However, the fractional representation of sources
of different ages is very model dependent (see below). 

     The sample plotted in figure 3 is complete for all Abell clusters  
within the volume $0.0268<z<0.09$ to a limiting flux density of $\rm 10~mJy$
at $1400~\rm MHz$ within 0.3 Abell radii.  
Because of the size of the primary beam of the VLA ($30\arcmin$ at 20cm), 
we are only complete to a search radius of $20\arcmin$ for sources brighter
than $\rm 10~mJy$.  .  
This sets a lower-limit of $z=0.0268$ for which we have uniformly surveyed
all clusters 
out to 0.3 Abell radii ($\sim 600~\rm kpc$ for $H_0=75$).   
Within this restricted volume, 
the distribution of sources in both 
radio and optical luminosity 
reflects a probability function for radio emission in rich clusters. 
Over this redshift range, the shape of the distributions will be 
governed by individual source evolution and not by cosmological evolution
in the radio source population ({\it e.g.}\ Meier \etal\ 1979). 
If we assume that the optical luminosity of the host galaxy does not change
appreciably over the lifetime of the radio source,
and that the radio power
evolves in time as the sources grow larger in size, the density of points
along the radio axis at a constant optical luminosity 
represents the amount of time radio sources spend in that state. 
We can therefore use these relationships to construct
bivariate radio luminosity functions for this sample (see \S\ 4.2)

\section{Luminosity Functions and Normalization}

     The univariate radio and bivariate radio/optical luminosity functions
will be derived from the complete sample of radio galaxies cataloged 
in Table 2 from Paper IV.  The univariate radio luminosity function
 will be compared to
that found by other investigators for non-cluster and field galaxies.  If
the dense environment of clusters influences radio source evolution
or formation, these effects should be seen as differences in the
shape of the radio luminosity function.  The bivariate functions
will be used to examine the importance of the optical luminosity of
the host galaxy in determining the radio properties.

     While several normalization schemes are commonly used,  
we have chosen to calculate the fractional
or differential radio luminosity function (\ie\ the fraction of galaxies
emitting in some radio power range normalized to the total number of galaxies
surveyed in each bin). 
For the bivariate case,
this also involves the determination of the number of galaxies which
could have been detected as a function of optical luminosity.  As a first
step, the bivariate luminosity functions were normalized to the number
of clusters surveyed (taking into account completeness in radio power; see
below).  These functions were first reported in Owen and Ledlow (1994).
To determine the number of galaxies which could have been detected, one must
either count the number of elliptical galaxies in each
cluster within
the Abell radius limit which were surveyed or could have been detected on the
radio maps and bin them according to their absolute magnitude (which is impractical),
or assume some distribution in number and optical luminosity for the elliptical galaxies
which can be scaled to the properties (richness) of each of the
surveyed clusters.  The differential bivariate luminosity function can then be
calculated
by computing the ratio :
\begin{equation}
\Phi(P,M) = {{n(P_{i},M_{j})}\over{N(P_{i},M_{j})}}
\end{equation}
where $\Phi(P,M)$ is the luminosity function as a function of radio power and
optical absolute magnitude, $n(P_{i},M_{j})$ is the number of galaxies detected
in the radio power bin {\it i} and magnitude bin {\it j}. and $N(P_{i},M_{j})$ is
is the total number of galaxies which could have been detected in bins {\it i}
and {\it j} (Auriemma \etal\ 1977).
For the univariate case, one
simply sums over all magnitudes in a given radio power bin.  In the bivariate
case, this ratio 
can be interpreted as the probability for a galaxy of magnitude $M\pm dM$ to be
radio emitting within a radio power $P\pm dP$.  The univariate function gives
the fraction of all galaxies which are radio
emitting over the range in the radio power bin.  The integrated luminosity functions
are found by forming a cumulative sum in equation (1) of the ratio 
of all galaxies with radio power 
$\geq P$ in each bin.  The integrated luminosity functions can also be
calculated seperately for each optical magnitude bin. 

     Because the sample is derived from a flux-limited survey, there will also
be a completeness correction which is a function of radio power.  
Using our flux
limit of $10~mJy$ and a redshift cutoff of $0.09$, the survey is complete
for $\log(P_{1400}) \geq 23.23$.  Below this power, radio galaxies could
not have been detected out to the maximum redshift of $0.09$.  In the radio
power bins below this value, the number of clusters which could have been
detected at a minimum flux density of $10~mJy$ is found by solving 
for the redshift limit ($z_{lim}$) at the midpoint
of the $P_{1400}$
bin.  The number of clusters surveyed with redshifts $\leq~z_{lim}$ is 
used to determine the number of galaxies surveyed, as described
in the next few paragraphs.

     The optical normalization function (or elliptical galaxy optical luminosity
function for clusters) was determined using data from
Dressler (1980).  Using photographic plates, Dressler catologued the positions
and morphological types for galaxies in 55 rich clusters.  Thirty-eight of
these were Abell clusters.  For twenty-eight of the Abell clusters, he also
included estimated V apparent magnitudes, bulge sizes, and ellipticities.  We 
have used his data for 25 of these clusters to calculate the elliptical galaxy
optical luminosity
function for rich clusters ($\Phi_{E}(M)$).
The other 3 clusters (Abell 0548, 1631, and 1736) are confused by superpositions
of two clusters, so were not included in the analysis.  
Using Dressler's galaxy classifications, we have determined
$\Phi_{E}(M)$
within a search radius of 0.3 corrected Abell radii. 
When 
Dressler has indicated an ambiguous Hubble type (\ie\/ S0/E, or Sa/0), we used 
the first classification given.  The apparent V-magnitudes were transformed
to R-Band colors using $V-R_{C}=0.59$ (Bessel, 1979), and converted
to absolute magnitudes using our adopted cosmology.  A K-correction was also
applied to the absolute magnitudes with the approximation : $K_{corr}=1.122*z$.

     For each cluster, the following procedure was taken : 1) determine the
projected distance from the cluster center for each galaxy as a fraction of
a corrected Abell radius, 2) reject all galaxies outside the nominal 
0.3 Abell radii, 
3) convert apparent V-magnitudes to absolute R-band magnitudes,
4) restrict galaxies to Hubble
classifications of {\sl E}, {\sl D}, or {\sl cD}, and  5) bin the galaxies in
units of 0.75 magnitudes over the range $-20.5$ to $-25.0$ and output the results.
Once these data were compiled
for all 25 clusters, the number of galaxies in each magnitude bin were
summed over
all clusters and tabulated.

     The procedure was to sum over the richness values for all
$25$ clusters, 
and scale the total number of galaxies in each magnitude bin to this value.
Our derived elliptical galaxy optical luminosity function is shown in figure 4.
This function can be compared to the type-specific optical
luminosity functions given in 
Binggeli \etal\ (1988) for the Virgo cluster. 
The magnitude scales can be 
compared by using a $B_{T}-R \sim 2.5$ (includes a correction for the cosmology).  
We verify that the two
functions have a similar shape.  By interpolating values from the plot in
figure 4 for the midpoint values of each magnitude bin, a
smooth-continuous normalization function can be found.
The result is the number of elliptical galaxies within 
the magnitude interval $M\pm dM$ 
within $0.3$ Abell radii of the cluster centers, scaled by Abell's galaxy counts. 
For the bivariate calculations, these percentages
are then multiplied by the $N_{G}$ from the total number of clusters which
could have been
detected in each radio power bin (taking into account the radio-completeness
correction).  
For the univariate case one simply sums the number of galaxies
over all magnitudes in each radio power interval.

     In most previous studies of radio galaxies, ellipticals and {\sl S0's} have
been examined together.  We have found very few ($\sim 3$) {\sl S0's} which
were radio emitting above our flux limit, and have eliminated these to examine
only the elliptical galaxy luminosity functions.
Because of the very small percentage of radio emitting {\sl S0's} in our sample,
normalizing to an {\sl E} + {\sl S0} luminosity function will change the
overall shape and the fraction of radio-loud galaxies in any given
luminosity range.  In order to compare our
work to previous studies, however, it is necessary to normalize our sample
 to the
combined {\sl E+S0} optical luminosity function.
An identical procedure to that described above was performed
using the data from Dressler, but including the {\sl S0's}.  This function
is shown in figure 5.  We see, as expected, the function is similar
for the brightest galaxies (most of which are ellipticals), but has a 
steeper faint-end slope when the {\sl S0's} are included.

\section{Luminosity Functions for the Complete Cluster Sample} 

\subsection{Univariate Radio Luminosity Function}

     The fractional univariate radio luminosity function (RLF) has been derived as discussed
in the previous section for the complete sample with $z<0.09$.  The binned data
in units of $(\#~detected)/(\#~surveyed)$
is listed in Table 1.  The ratios are the fraction of galaxies over all
magnitudes brighter than -20.5 which are radio emitting in the radio power range of
each bin.  This function is shown in figure 6.  The error bars
are calculated as $n^{-\frac12}$ sampling errors (percentage uncertainties). 

     We see that the function is relatively flat for radio powers up to
the break power ($\sim 10^{24.8}$), and falls off radidly with increasing
$P_{1400}$.  The slopes on each side of the break are : $-0.11\pm0.05$ for
$P_{1400}<10^{24.8}$ and $-1.52\pm0.33$ for $P_{1400}>10^{24.8}$.  The
fit above the break is obviously more uncertain with only three data points,
the last one being an upper limit.  As compared to Sadler \etal\ (1989), whose
survey extends to $10^{20}~W~Hz^{-1}$ (converted from $5~GHz$ to $1400~MHz$ using
$\alpha=0.75$), our faint-end slope is consistent with a gradual rise in the
fraction of detected galaxies with decreasing radio power.  This is expected
at lower radio powers as more and more galaxies are detected at small levels
of radio activity and the contribution from spirals and {\sl S0's} begins to
increase.  de Ruiter \etal\ (1990), for a sample of {\sl B2} radio galaxies,
found a faint-end slope of $-0.3\pm0.1$ and above the break, $-1.3\pm0.1$.
His normalization is different, however, as he was forced, by
the definition of the sample, to normalize to galaxies per unit volume.

     From examination of figure 3, we can determine the origin of
the break power in the univariate luminosity function.  The univariate function is
found by summing over all optical luminosities.  Thus, if one collapses
the data in figure 3 along the x-axis, and were to examine the
distribution of sources in bins of radio power, the sources above the
break originate entirely from the group of higher-power sources at bright
optical luminosities.  Since the {\sl FR I/II} division has a slope in this plane
$\propto L_{opt}^{1.8}$, the width of the break in the luminosity function is a result
of projecting this sharp division onto a single axis, leaving radio power as
the only variable.  We should also notice that becuase of the very low number of
{\sl FR II's} in the sample (3), all the objects above this break power are {\sl FR I's}
from the highest radio power and optical luminosity bin.  All the {\sl FR II's}
show up in the $24.6$ bin.  Therefore, the break in the univariate RLF
corresponds only approximately to the {\sl FR I/II} division.

     As mentioned in \S\ 2, our sample was selected to examine
a sample of radio galaxies unaffected by cosmological evolution, but rather
to examine a large number of radio sources in different stages of their
individual evolution.  A previous study by Meier \etal\ (1979) of the radio
luminosity function of {\sl B2} and {\sl 3C} objects, found no evidence of
any difference in the luminosity functions for objects with redshifts $<0.12$.
Thus the shape of the luminosity function is most dependent on initial 
conditions in the radio source and 
individual radio source evolution rather than any effects due to epoch.
The break power, which is coincidently in the range where the transition from
{\sl FR I} to {\sl FR II} morphology occurs, is most likely a result of
the different lifetimes for sources as a function of radio power.

    This last point is very important to our understanding of radio galaxy
evolution.  While the luminosity functions describe the existing population
of radio sources at any given epoch and radio power, it is most useful
as a constraint to models of radio source evolution. 
Meier \etal\ (1979) presented several models to explain the shape and
position of the break in the RLF.  As we will examine in the next few
sections, the suggestion is that the RLF is not dependent on
environment, and at least to redshifts $\sim 0.2$, is not largely
dependent on epoch.  Therefore, the shape of the RLF must represent
some intrinsic property of radio galaxies. If radio sources have very
long lifetimes ($\sim$ a Hubble time), the shape of the RLF must
reflect the initial RLF at the time the sources were created.  If
radio sources have short lifetimes (in a continuous cycle of dying and
new sources created) the RLF
is dependent on the initial RLF and the lifetimes of the sources
as a function of radio power.  In this scenario, the amplitude of the
function and the location of the break can evolve in time.  
It is therefore very interesting to
determine if the break in the RLF corresponds to the {\sl FR I/II}
division from figure 3.

     In paper V, we suggested that radio sources might be a transient
phenomenon which occurs in all ellipticals at some time (or many times)
over the course of their lifetime.  We can use the univariate luminosity
function listed in Table 1 as a constraint to test this idea.  If 
we integrate the univariate RLF, we find that 14\% of ellipticals 
with $M_{R}$ brighter than -20.5 are detected with $P_{1400}\geq 10^{22}~
\rm Watts~Hz^{-1}$ within our search volume.  Using an age 
of $10^{10}~\rm years$ for a typical elliptical galaxy, 
if radio source lifetimes are $\leq 1.4\times 10^9~\rm years$, our RLF is 
consistent with the idea that all ellipticals may at some time have powerful radio 
sources.  
Typical lifetimes from spectral aging arguments, light-travel time
calculations and models (Eilek and Shore, 1989) suggest that 
$1 \times 10^{8-9}~\rm years$
may be typical for radio sources.  Thus, the suggestion that the parent
population of radio sources includes {\bf all} ellipticals may be consistent
with the observed statistics.

\subsection{Bivariate Radio/Optical Luminosity Function}

     In this section, we divide the detected sources into bins of both optical 
luminosity and radio power 
to investigate the dependence of the shape of the luminosity
function on the optical luminosity of the host galaxy.  A comparison will be
made to samples of objects not chosen to reside in clusters to examine the
effect of the global environment on the formation and evolution
of radio galaxies.

     In Table 2, we show the results of binning the data into $0.75$ magnitude
intervals in $M_{24.5}$ and $0.4$ bins in $\log P_{1400}$.  The normalization
function is that derived in \S\ 3 (figure 4) for
only elliptical galaxies.  The bivariate luminosity functions are shown in 
figures
7A-D.  We note that 1) all of the functions are
relatively flat for powers less than $10^{25}$ (except in the faintest magnitude bin),
 2) the fraction of galaxies
which were detected in bins of $\log P_{1400}$ increases over the first
two magnitude bins ($-21.62$,$-22.38$), but is about the same in the upper
two bins ($-23.12$,$-23.88$).  This point will be discussed in the next paragraph.
A K-S test between successive magnitude bins (for $P_{1400}<10^{25}$) suggests
that the shape of the
individual functions is not strongly dependent on optical luminosity. 
This supports the contention that the radio and optical luminosities of individual 
galaxies are not strongly correlated. However,
the location of the break power does appear to shift by $\sim 0.4 \log P$ per
$0.75$ magnitude interval.  
In the last bin ($-23.88$), the function is consistent with having not reached the
break power up to $10^{26}$.  According to the fit from figure 1 for the
{\sl FR I/II} division, at this luminosity, the break might occur at a radio
power of $\sim 10^{26.1}~W~Hz^{-1}$.  We thus suggest that the break power in
the bivariate RLF corresponds to the {\sl FR I/II}
division shown in figure 3, and is a function of optical luminosity. 
For the univariate function, averaging over all magnitudes, the break results
from projecting the {\sl FR I/II} division onto only one variable and only for
the fainter galaxies does it correspond to the actual {\sl FR I/II} division
in radio power.

  The integrated bivariate luminosity functions are plotted in figure 8. 
From these
plots, we can directly measure the probability for a galaxy with absolute 
magnitude in the range $\rm M \pm dM$ to have a radio source with radio power
$\rm P_{1400} > P$.  
Objects with $\rm M=-22.4\pm0.37$ are three times more likely to have
radio sources than objects 0.75 magnitudes fainter at nearly all radio
powers.  Galaxies brighter than $\rm M=-22.4$ are about 2.5 times more
likely to be radio emitting. 
However, the 
brightest two bins are consistent with having equal probabilities. 
While the normalization
is the most uncertain in this range of optical luminosity because the optical
luminosity function is changing rapidly, the most likely effect from our
normalization is that we slightly underestimated the number of galaxies in the last bin.
The median magnitude of the objects in this bin is $-23.72$, which would
change the normalization only slightly in the direction of a lower detection
ratio (see figure 4), not increase it. 
The median magnitude in the $-23.12$ bin is $-23.10$. The normalization is
consistent with the actual mean values of the data within a few percent of the bin
width. 

     Therefore, galaxies brighter than $M_{24.5}=-22.7$
may be equally likely to have radio sources with radio powers $>10^{22}$.
However, Burns (1990) found that cD galaxies typically have 
steeper radio spectral indices than typical of most
radio sources.  This result is particularly strong for 
cD's with compact or amorphous radio emission possibly associated with 
a cooling flow.  
Therefore, individual objects in the brightest optical magnitude bin may
have a large range in spectral index which may bias the average radio power
level to lower values.  However, on the log-log plot, it would require a large 
increase in 20cm flux-density for a majority of the sources to substantially
separate the two curves. It seems likely that spectral-index
variations may contribute to the observation, but it is far from clear that 
this alone accounts for the similarity in the amplitudes for the two magnitude bins. 
Thus the observed narrowing separation between
the functions for increasing optical luminosity may be a real effect. 

     Also from figure 8, we see that the break power in the integral
functions shifts with optical luminosity as well. 
This supports the idea that the radio
power depends only weakly on optical luminosity, and that this dependence is
directly related to the maximum power an {\sl FR I} can take at any given
optical luminosity. 
These effects would be difficult to see 
with larger magnitude bins. 
With larger bins, the numbers of sources in the brightest bins 
would cluster towards the faint end because of the shape of the 
optical luminosity
function.  This effect would skew the probabilities towards those 
of fainter magnitudes
and smear out the distributions in figure 6. 

     Given the optical dependence of the FR I/II break and the 
optical dependence of the luminosity functions discussed above,
it seems likely that the host galaxy properties directly affect
both the initial conditions in the radio source as well as 
the subsequent evolution and lifetime of the sources.  Following 
the analysis of Cavaliere \etal\ 1971, Eilek (1993) performed
analytic calculations for the radio power and size evolution of
FR I's propogating through a constant density medium.  The luminosity
function can be represented as :
\begin{equation}
{{\partial{N}\over\partial{t}} + {\partial{}\over\partial{P_\nu}}(N~{dP_{\nu}\over{dt}})} = 
S(P_{\nu},t)
\end{equation}
where N is the source counts as a function of total monochromatic 
radio power over a time-interval 
$dt$, ${dP_{\nu}/dt}$ is the change in monochromatic power with time
(incorporating spectral aging), and $S(P_{\nu},t)$ is the source function
for the number of sources created at power $P_{\nu}$ over a time $dt$.
Solutions for $N(P_{\nu},t)$ show that the radio power increases up to 
a time $t_{sy} \propto (B^{-3/2} \nu^{-1/2})$ (the synchrotron lifetime 
as a function of the energy of the radiating electrons and the magnetic field
strength), and thereafter
decreases as the source lobes come into pressure balance with the ambient
medium and the electrons simply radiate on a timescale $t_{sy}$.  This 
analysis nicely reproduces the slope in the univariate RLF below $10^{24}~
W~Hz^{-1}$, but predicts a very slow decay in the source power.  This 
would predict many more old steep-spectrum sources than is observed.
Thus, the sources must fade on much shorter time scales (Eilek, 1993). 
{\it In situ} 
particle acceleration via either turbulance in the lobes or 
Fermi accleration does not significantly affect the time-decay of the
radio power averaged over the entire source (Eilek and Shore, 1989)
as these processes operate on small size scales.  

     This analysis implicitly assumes that the ambient atmosphere of all 
sources is the same, effectively ignoring possible dependencies on 
the host galaxy properties.  From figures 7 and 8, it is clear that 
$N(P_\nu,t)$ must also be a function of $M_{opt}$.  However, Eilek (1993) 
found that the choice of ambient density or pressure had much less 
effect on the evolution as compared to the initial spread of 
intrinsic beam powers
($P_{jet}$).  
The source 
counts for the brightest galaxies may be consistent with a spread of initial
beam powers which is skewed to higher values and an extensive ISM which allows 
the source to grow to a larger linear size (over a longer time span, assuming
the jet velocity is similar for all FR I's) 
before reaching a contact pressure 
discontinuity at the interface between the galaxy ISM and the ICM. 
For $P<10^{25}$ at a constant radio power, sources at $M=-23$ may have 
lifetimes between 3 and 15 times longer than sources in galaxies 
a magnitude fainter.  In another interpretation, if all radio galaxies 
have similar lifetimes, the difference between galaxies of different
optical luminosities may be due to the spread in intrinsic beam powers
plus a functional form for $dP_\nu/dt$ which depends on the properties
of the galaxy ISM.  
Modification of equation (2) to include the 
dependence on $M_{opt}$ and solutions for $N(P_{\nu},M_{opt},t)$
constrained by our observed luminosity functions
should allow one to sort out the dependencies between $P_{jet}$, $dP_{\nu}/dt$,
and the luminosity and size of the host galaxy. 
Clearly these bivariate 
luminosity functions will provide an important constraint to understanding
FR I evolution.  Some of these ideas have already been incorporated into a model
for the optical dependence of the FR I/II break by Bicknell (1995). 

\section{The Bivariate Luminosity Function in Different Environments}

\subsection{Comparison to data from Auriemma et al. (1977)} 

     One of the first determinations of the local bivariate luminosity functions
was made by Auriemma \etal\ (1977).  Auriemma \etal\ constructed their sample from
surveys of the {\sl B2} and {\sl 3CR} samples (Colla \etal\ 1975), a survey
of {\sl E} and {\sl S0} galaxies not classified as cluster objects,  
and data from five rich clusters (Jaffe and Perola, 1976).
The sample included $128$ objects not-selected to reside in clusters, and $17$
detections from rich clusters.  Using the argument that the univariate
RLF was not significantly different in and out
of clusters, the samples were combined to produce a single luminosity function. 
Auriemma \etal\ does not rule out, however,
that the fraction of radio galaxies might be different up to a factor of
two between the two environments.  Thus we should be able to
compare our sample (of only rich cluster radio galaxies) to their combined group
to examine what effect, if any, the local galaxy density has on the
radio galaxies.  The average density of galaxies in rich clusters is of the
order $100~Mpc^{-3}$ as compared to $<10~Mpc^{-3}$ in poor galaxy clusters or
the field. 

     In order to construct the fractional luminosity functions, 
Auriemma
normalized the binned values to an estimation of the
optical luminosity function per unit volume.  This function approximates
the average density of galaxies of types {\sl E} and {\sl S0} within
the search volume.  This correction only applied to the {\sl B2} and {\sl 3CR}
samples.  For the rich-cluster observations, knowledge of the fractional
functions were obtained directly.  As they point out in their paper, their
may be a bias introduced by choosing an unknown function for this purpose.
In addition, their function does not have the expected shape based on our
results in figure 4. 

     We have compared the normalization function from 
Auriemma \etal\ to a complete sample
of galaxies chosen from the {\sl UGC} catalog and surveyed at Arecibo
by Dressel and Condon (1978).  From this survey of over $1600$ galaxies,
the volume density of E and S0 type galaxies should be
similar to that covered by the B2 and 3CR surveys.
Franceschini \etal\ (1988) computed the optical luminosity function
in bins of one magnitude intervals measured on the Zwicky magnitude system. 
A comparison of the Franceschini \etal\ and Auriemma \etal\ functions
are shown in figure 9.  We have corrected the cosmology of
both data-sets (both magnitude and volume corrections) to $H_{0}=75$.
We see that for the faintest and brightest
bins, the two functions are in reasonable agreement.  For the magnitude
range $-19.5 < M_{pg} < -21.5$, however, Auriemma's function predicts over
a factor of two less objects in the search volume.  Fransechini's data
is much closer to our derived function and exhibits the expected break in the
optical luminosity function which is absent from Auriemma's approximation.
We have therefore renormalized the Auriemma \etal\ data to these revised
galaxy counts in the two affected magnitude bins.  Conversion to our $M_{24.5}(R)$
system was made using $M_{pg}-M_{R}=1.74$ (using $M_{pg}-M_{V}=1.15$ from 
Auriemma \etal\ 1977; and $M_{V}-M_{R_C}=0.59$; Bessel, 1979) 
and have rebinned our
data to match that of Auriemma \etal\ 
(one-magnitude bins centered on $M_{R}=-21.86,-22.86$,
and $-23.86$).

     Our bivariate functions and those from Auriemma \etal\ (renormalized for
the two middle bins) are plotted in figures 10 A,B, \& C. 
The scales are identical for each magnitude range to allow easy comparison
over the range in optical luminosity. 
There are only $4$ detections in the $-20.86 < M < -21.86$ interval, so
this function was not plotted.  In general, we see quite good agreement
between the datasets.  As before, the errorbars are $n^{-1/2}$
uncertainties.  In many cases,
especially in the Auriemma \etal\ data, 
there were no detections in a bin.  In these
instances, upper limits are plotted with the value of one detection.
For radio powers $\leq 10^{24.5}$, visually the data agree very well.  We have
very few detections above this power in any of the bins, whereas Auriemma \etal\
supplemented their sample with the 3C objects which mostly fall
above this power.  We therefore will only compare our samples over the radio
power range $22 \leq \log P_{1400} \leq 24.5$.

     To compare the shapes of our functions to Auriemma's, we have calculated
linear-regression fits for $P<10^{24.5}$.  Auriemma chose this power cutoff
as the approximate position of the break ($P^{*}$), and calculated fits above and 
below $P^{*}$ for the slope. We follow an identical analysis, but keeping
in mind that $P^{*}$ is actually a function of optical luminosity. 
In this radio power interval,
the following fits were derived for the differential functions (of the form
$F~d\log P  =  B~P^{A}~d\log P$ : 

\begin{mathletters}
\begin{eqnarray}
M=-21.86~~~~~F~d\log P & = & +3.87(\pm1.9)P^{-0.26\pm0.08}~d\log P \\
M=-22.86~~~~~F~d\log P & = & -0.63(\pm1.6)P^{-0.05\pm0.07}~d\log P \\
M=-23.86~~~~~F~d\log P & = & -1.35(\pm1.9)P^{-0.003\pm+0.08}~d\log P
\end{eqnarray}
\end{mathletters}
We therefore see that the functions below $P^{*}$ become
flatter with increasing optical luminosity.
Also, the functions are flat out to higher radio powers as $L_{opt}$ increases.
From figure 3, this is consistent with the position
of the {\sl FR I/II} division.  In the lowest magnitude bin, the break
occurs at $\sim 10^{24.5}$.  In the next magnitude bin, the division
lies at $\sim 10^{25.4}$.  So by choosing a constant $P^{*}$, as Auriemma
and we have done in determining the slopes of the individual functions,
the expected effect from figure 3 is that at bright magnitudes,
there are still many sources with powers 
$\geq P^{*}$; hence the nearly $0$ slope.
In the $-21.86$ magnitude bin, the number of sources is already dropping
off very rapidly by $10^{24}$, so the calculated slope over this range
is much steeper.
This fact confirms our suggestion that the position of the break $P^{*}$ is
a function of optical luminosity and coincides with the {\sl FR I/II} division.
We have insufficient detections above
powers of $10^{25}$ to determine the slope above $P^{*}$.  Auriemma found
that the slope was independent of optical luminosity at higher radio powers.

     In order to quantitatively determine whether the bivariate luminosity
functions from our sample are consistent with those from Auriemma \etal\, 
we have
performed several statistical tests.  Because we are mostly interested
in whether or not the fraction of radio galaxies detected (a measure
of the probability for a galaxy to be radio emitting) is equivalent
between samples from the different environments, tests which examine
the mean values of the distributions are most appropriate.  We have used
two tests for this purpose; a parametric {\sl T-test} and
a non-parametric Wilcoxon rank test (WR).  We also calculate the Spearman's
$\rho$ (SR) correlaton coefficient from the derived ranks.  
Spearman's $\rho$ is a measure of the signficiance of the correlation
between the two samples.  For negative values, the samples are
anti-correlated. Positive values indicate a correlation.  
See Conover (1980) for a more detailed description of these statistics. 

     The results of the statistical tests are summarized in Table 3. 
In all cases, the signficance levels are interpreted in the sense that
small percentages indicate significant differences in the means.  
In none of the tests are the distributions different at levels higher than
3\% (which is the
value from the WR test for the $-21.86$ magnitude bin). 
Also, no two tests on the same set of functions show significant evidence 
for differences in the parent populations. 
Within
statistical uncertainties, the samples are consistent with having been
drawn from the same parent distribution.  Interestingly, this result was 
initially suggested 
by Fanti (1984) from a sample of clusters about one-fourth the size of 
our current sample.  

     Sources of additional errors in the determination of the
luminosity functions were discussed by Auriemma \etal\  
We have already corrected for the underestimation in the number density
of galaxies in two magnitude intervals as discussed above. 
Most likely the largest
source of error derives from the optical magnitudes from their sample.
We estimate (see Ledlow and Owen, 1995b) that our isophotal magnitudes are accurate
to $\sim 0.05~\rm magnitudes$.  
The Zwicky magnitudes measured for the {\sl B2} and {\sl 3CR} samples, however,
show systematic biases $\geq 0.3$ magnitudes (Colla \etal\ 1975; 
Giovanelli and Haynes, 1984).  We also expect some error in transforming $M_{pg}$ to our
magnitude system.

     Probably the most surprising observation from figure 10 
is the offset in the detected fraction in the brightest magnitude bin. 
Because ellipticals brighter than $-23$ are mostly found in
clusters, we would expect that the detections in this bin are dominated
by the cluster objects present in Auriemma's sample.  We would therefore
expect that the functions in this bin would have the closest agreement.
Of course, if the probability for
radio emission does not change with environment, the fraction detected
would simply scale with number of galaxies surveyed, as we have apparently
observed (see also Ledlow and Owen, 1995a).  
A possible explanation for less-than-perfect agreement in the
$-23.86$ bin may be a result of the errors in the Zwicky magnitudes.  If the
magnitudes are uncertain to within $\geq0.3$, many more objects from the
$-22.86$ bin will be scattered into the $-23.86$ range because of the
shape of the optical luminosity function.  From figure 5, there
are approximately $4$ times as many galaxies in the $-22.86$ bin than
in the next brightest bin.  Therefore, while the normalization function
is constant, the number of detections will be artificially higher in
the $-23.86$ bin simply because of the error in the magnitudes and
the number statistics.  This would produce the observed higher detection
rate seen in figure 10. 

     We have recalculated the integrated bivariate luminosity functions
for our cluster sample.  These functions are
shown in figure 11.  We see that, (as also found by Auriemma), the
fraction of galaxies at all radio powers $>10^{22}$ is greater for
brighter galaxies.  
From figure 8, we found that brighter than $M=-22.7$, 
the probability for a galaxy to be radio emitting above
this flux limit was constant.  This fact can be understood simply
by the effect of the bin width.  As one increases the size of the
magnitude bin, more and more galaxies of fainter luminosities are
averaged because of the shape of the optical luminosity function.  This
will tend to lower the detection rate if the probability lowers with
decreasing luminosity, as is observed.  By reducing the magnitude bin size
to smaller and smaller values, the resolution increases, thus enabling
a more accurate representation of the optical dependence.
It is because of our large sample
size that we have been able to see the non-linear dependence between the 
number density of sources at a given radio power and the optical luminosity
(which probably depends on the radio power evolution of the sources in 
different host galaxies). 

These
findings suggest that either the local galaxy density and large-scale
environment in which the radio galaxy lives is not important to the
formation and evolution of radio sources, or that the environments
between these samples are not that different.

\subsection {Comparison to the Sample from Sadler \etal\ (1989)}

      Sadler \etal\ (1989) observed a sample of $248$ elliptical and
{\sl S0} galaxies selected from the southern {\sl ESO/Uppsala Catalogue} 
and surveyed at $2.7$ and $5~GHz$ with
the Parkes single
dish telescope (Sadler, 1984a,b,c).  All sources from this sample
north of $\delta = -45^{\circ}$ were reobserved with the VLA to map
the extended structure and extend the detection limit to lower
radio powers.  None of these objects were selected to reside in
clusters.  Thus, this sample provides another good comparison
with which to test the effect of environment.  In addition,
the flux limit of this survey extends two orders of magnitude
lower than ours ($\sim 10^{20}~W~Hz^{-1}$ shifted to $1400~MHz$
assuming $\alpha = 0.75$, and $F_\nu \propto \nu^{-\alpha}$).

    Sadler \etal\ discovered that nearly all ellipticals (and {\sl S0's}) with
magnitudes brighter than $\sim -20.3$ are detected at radio powers 
$\geq 10^{20}$.
Galaxies fainter than $-20.3$ are much weaker in the radio, and thermal
emission from {\sl HII} regions, or non-thermal radio emission from supernovae
may be the origin of the emission.
This is consistent with our results.  From our original survey 
(Ledlow and Owen, 1995a)
all galaxies with magnitudes fainter than -21 which were identified with radio
sources $>10~mJy$ were found to be background sources. 
Because of the slope of the {\sl FR I/II} division in
figure 3, the maximum power of an {\sl FR I} with magnitude $\sim -20$
is a factor of $10$ smaller ($10^{23.5}$) than for the faintest galaxy in our sample.
Since the probability for a radio galaxy to be radio emitting above $10^{22}$
is falling off rapidly in this optical luminosity range
(figure 8), it is not surprising that we detect no sources in our
sample at these optical luminosities. Sadler's results suggest that all {\sl E}
galaxies are active at some radio power.  Thus, the definition of {\lq\lq}radio-quiet\rq\rq
is very dependent on the flux-limit of the survey.  
Ledlow and Owen (1995b) found that cluster 
radio galaxies could not be distinguished optically from randomally selected
cluster galaxies. These observations suggest two possible interpretations;
1) all galaxies of this type have central radio sources at some power
level, and it is the properties of the interstellar medium of the galaxy
and the dynamics near the central engine which determine the power and
size of the radio source; and 2) assuming the lifetime of radio sources
is short compared to a Hubble time, or the lifetime of the parent galaxy,
all elliptical galaxies have powerful radio galaxies at some time in their
evolution (perhaps several times). 
These two ideas are not mutually
exclusive.  It seems likely that the properties of the interstellar
medium (and the size of the galaxy) may be important in both cases. However,
it is unclear whether some galaxies never have high power extended radio sources
because of (1) or whether we are just observing these galaxies in a
temporary quiescent state.

     Sadler \etal\ calculated bivariate luminosity functions binned in 
$1$ magnitude
intervals and $1$ dex wide in $\log P$.  All her observations are reported
at $5~GHz$.  We transformed the $5~GHz$ powers to $1.4~GHz$ assuming 
$\alpha=0.75$. 
The magnitudes reported in her paper are in the $B$ system, and were transformed
identically as done for the Auriemma \etal\ (1977) data.  After these
transformations, the radio power range of her sample ranges from $10^{20.9-24.4}$
at $1.4~GHz$ in $1$ magnitude bins centered on $-21.36,-22.36,$ and $-23.36$.  
We have rebinned our cluster sample identically in order to allow a direct comparison.
Our functions are 
overplotted with those of Sadler \etal\ in figure 12.  Because
the faintest galaxy in our sample is $-21.25$, the faintest bin
is not shown.  All of our objects would have fallen in the upper half of this
bin where Sadler's objects were more evenly distributed.  

     Because of the large bins in radio power, only two or three
points along the function actually overlap our sample.  
Both functions are consistent within the errorbars in the overlapping regions.  
For lower radio powers, Sadler's function continues to rise whereas ours
remains flat in our last data point.  However, we have very poor statistics 
in this bin (only one detection).  
Both functions are 
in agreement and consistent with near zero slope.  A K-S test
for both magnitude intervals rules out a significant deviation beyond the 
10\% level.  We therefore conclude that there is no significant difference
in the fraction of detected galaxies between the two samples.  
Elliptical galaxies in clusters
appear to have equal probabilities of being detected as radio loud 
as compared to elliptical galaxies chosen from random samples.  In addition,
elliptical galaxies over the range in optical luminosities 
from our sample 
have equal, or even slighly higher, probabilities of containing radio sources
nearly twenty times fainter, which we would have presumably detected given 
a lower flux limit. 
This confirms the trend from figures 2 \& 3, 
that while bright galaxies are more likely to contain more powerful radio 
sources than fainter galaxies (because of the {\sl FR I/II} break), they are
nearly equally likely to have faint radio emission. 
Most ellipticals are radio emitting at some power level regardless of 
environment.  Thus, the optical dependence of the FR I/II break appears 
to be consistent 
for objects in a variety of environments. 
Given the errors in 
the Zwicky magnitudes, and errors induced by transforming to our magnitude
system,  
however, we cannot rule out that the slope and 
intercept of this relation may be different in and out of 
rich clusters.  We will address this issue in a future paper. 
     
     All these results suggest that either the properties of the environment 
in the centers of rich clusters does not affect the
radio source population and evolution or that the environment is not
significantly different inside and outside catalogued rich clusters.
The primary parameter which influences the radio properties and probability 
of detection is the optical
luminosity of the host galaxy.

\section{Conclusions}

     We have shown that the {\sl FR I/II} division is a function of
optical luminosity as well as radio luminosity.  However, there is no
strong correlation between the radio and optical luminosities of
individual galaxies.  Along a projection in the radio power plane for
a constant optical luminosity the {\sl FR I} galaxies can take on a large range
in radio powers from the weakest detectable powers in our survey up
to powers on the order of the {\sl FR I/II} break luminosity.  With
increasing optical luminosity, the {\sl FR I/II} break power increases
as $L_{opt}^{2}$.  There are, however, {\sl FR II} sources below this break
and {\sl FR I's} above the break.  These features suggest that the two classes
are not as distinct as once thought.  It is hard to imagine from these
diagrams that there is not some connection between the two classes. 

     The univariate RLF was calculated for our cluster sample and compared
to published results from samples with very different selection criteria.
Both the slope (shape) and the integrated fraction of galaxies as a function
of radio power appear to be invariant quantities; not dependent on the
environment in which the radio galaxies live.  Therefore, while the
morphology and global properties (\ie size) of radio sources in and out of rich
clusters may differ,  the evolution in terms of the radio power and
lifetime do not change with environment.  The slope of the RLF below
the break is a slowly rising function (slope = $-0.33$) which continues
to rise to at least two orders of magnitude fainter in radio power than our
VLA cluster survey (Sadler \etal\ 1989).
The position of the break
in the univariate RLF only roughly corresponds to the {\sl FR I/II} break.
This results from averaging over optical luminosity and projecting a
two-dimensional function onto a plane with a single variable.

     The bivariate RLF was calculated in bins of $0.75$ magnitudes and
$0.4$ in $\log P_{1400}$.  It was found that the slope of the functions
for $P<P^{*}$ are dependent on optical luminosity.  With increasing
optical luminosity the RLF flattens as more sources are detected
at higher radio powers.  
All of these properties can be related to the optical dependence of the
{\sl FR I/II} break which coincides with the break in the bivariate
luminosity functions.  Therefore, $P^{*}$ is also a function of $L_{opt}$
and has the relation : $P^{*} \propto L_{opt}^{2}$.

     From the integral bivariate RLF's, we have shown that the probability
for a galaxy to have a radio source with $P>10^{22}~W~Hz^{-1}$ increases
with optical luminosity up to a magnitude $\sim M^{*}-0.5$.  While there
is a difference of $\sim 2.5$ magnitudes between $M^{*}$ and the brightest
galaxy in the cluster, the probability of detection is essentially constant
in this luminosity interval.  Thus the evolutionary track of $dP_\nu/dt$
for FR I's is a fairly strong function of the optical luminosity (or other
properties which are related to it), however this dependence tends to 
weaken with increasing optical luminosity. 

    Statistical tests between our cluster luminosity functions and those
derived from samples not selected to be in clusters indicate no significant
dependence on the local galaxy density.  This result is consistent with
the invariance of the detected fraction of radio sources as a function
of richness (Ledlow and Owen, 1995a). 
Both the shape of the luminosity functions
as well as the total fraction of detected sources appear invariant with
respect to the environment.  This suggests that the luminosity function
is representative of the intrinsic properties and evolution of radio galaxies
as a class. 
The shape of the luminosity function
is directly related to individual radio source lifetimes and is
a function of both the
radio power and the optical luminosity of the galaxy.

\acknowledgements

M.J.L. acknowledges Jack Burns, John Stocke, and Jean Eilek for helpful 
conversations on this subject, and NRAO for financial support in the form
of a predoctoral fellowship during which much of this work was completed.
We also thank the referee, Chris O'Dea, for helpful comments on the text.

\newpage

\begin{deluxetable}{ccr}
\tablewidth{280pt}
\tablenum{1}
\tablecaption{Univariate Radio Luminosity Function} 
\tablehead{
\colhead{$\log~P_{1400}~(min)$}   & \colhead{$\log~P_{1400}~(max)$} & 
\colhead{Fraction}}
\startdata
22.028&22.428&2/59 \nl
22.428&22.828&7/334  \nl
22.828&23.228&26/1398 \nl
23.228&23.628&41/2210 \nl
23.628&24.028&29/2210 \nl
24.028&24.428&30/2210 \nl
24.428&24.828&31/2210 \nl
24.828&25.228&14/2210 \nl
25.228&25.628&4/2210 \nl
\enddata
\end{deluxetable}

\clearpage

\begin{deluxetable}{crrrrc}
\tablewidth{360pt}
\tablenum{2}
\tablecaption{Bivariate Radio/Optical Luminosity Function}
\tablehead{
\colhead{$\downarrow~\log~P_{1400}$}  & \colhead{$-21.62$} & 
\colhead{$-22.38$}          & \colhead{$-23.12$}          &
\colhead{$-23.88$}         & \colhead{$\leftarrow~M_{24.5}$}} 
\startdata
22.23&1/18&0/14&1/7&0/2& \nl
22.63&1/104&1/79&5/41&0/14& \nl
23.03&8/434&9/331&5/173&4/60& \nl
23.43&11/686&12/523&12/274&6/95& \nl
23.83&6/686&14/523&6/274&3/95& \nl
24.23&4/686&14/523&11/274&1/95& \nl
24.63&3/686&8/523&16/274&2/95& \nl
25.03&0/686&2/523&5/274&6/95& \nl
25.43&0/686&0/523&2/274&1/95& \nl
\enddata
\end{deluxetable}

\clearpage

\begin{deluxetable}{crcrcc}
\tablewidth{450pt}
\tablenum{3}
\tablecaption{Statistical Tests Between Samples}
\tablehead{
\colhead{$M_{24.5}$}        & \colhead{T}                  &
\colhead{Significance}      & \colhead{$\rho$}             &
\colhead{Significance}      & \colhead{WR Significance}}
\startdata
$-20.36 \leq M < -21.36$&-2.08&0.06&0.57&0.18&0.03 \nl
$-21.36 \leq M < -22.36$&-0.61&0.55&0.70&0.04&0.49 \nl
$-22.36 \leq M < -23.36$&0.52&0.61&0.37&0.41&0.08 \nl
\enddata
\end{deluxetable}

\clearpage

\begin{figure}[p]
\plotone{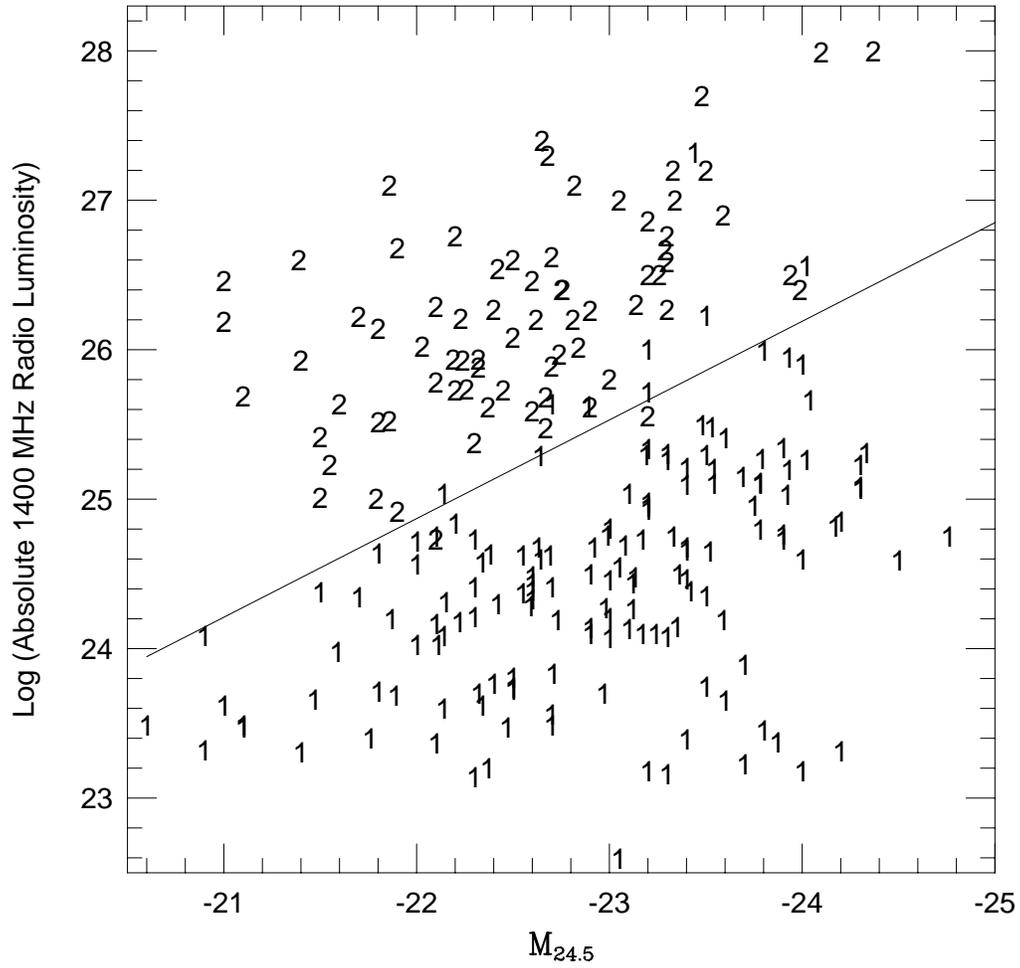}
\caption{FR I/II diagram for a sample of objects selected from the 
literature (see text for references).  Axes are the logarithm of the 
absolute radio power at $1400~MHz$ and the absolute isophotal magnitude
of the galaxy measured to $24.5~\rm magnitudes~arcsec^{-2}$ in the 
rest-frame of the galaxy.  Reprinted from Owen and Ledlow, 1994.}
\end{figure}

\clearpage

\begin{figure}[p]
\plotone{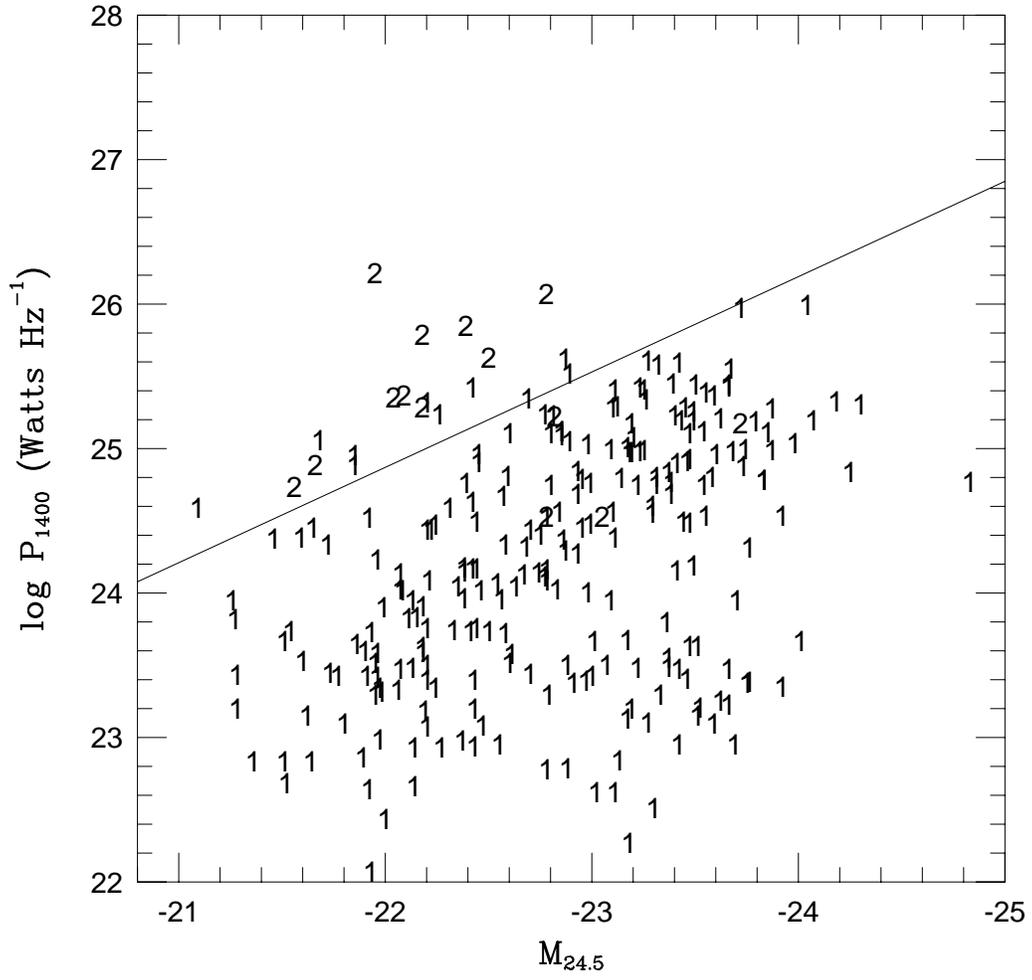}
\caption{FR I/II diagram for our entire cluster sample from $0.026<z<0.25$.}
\end{figure}

\clearpage

\begin{figure}[p]
\plotone{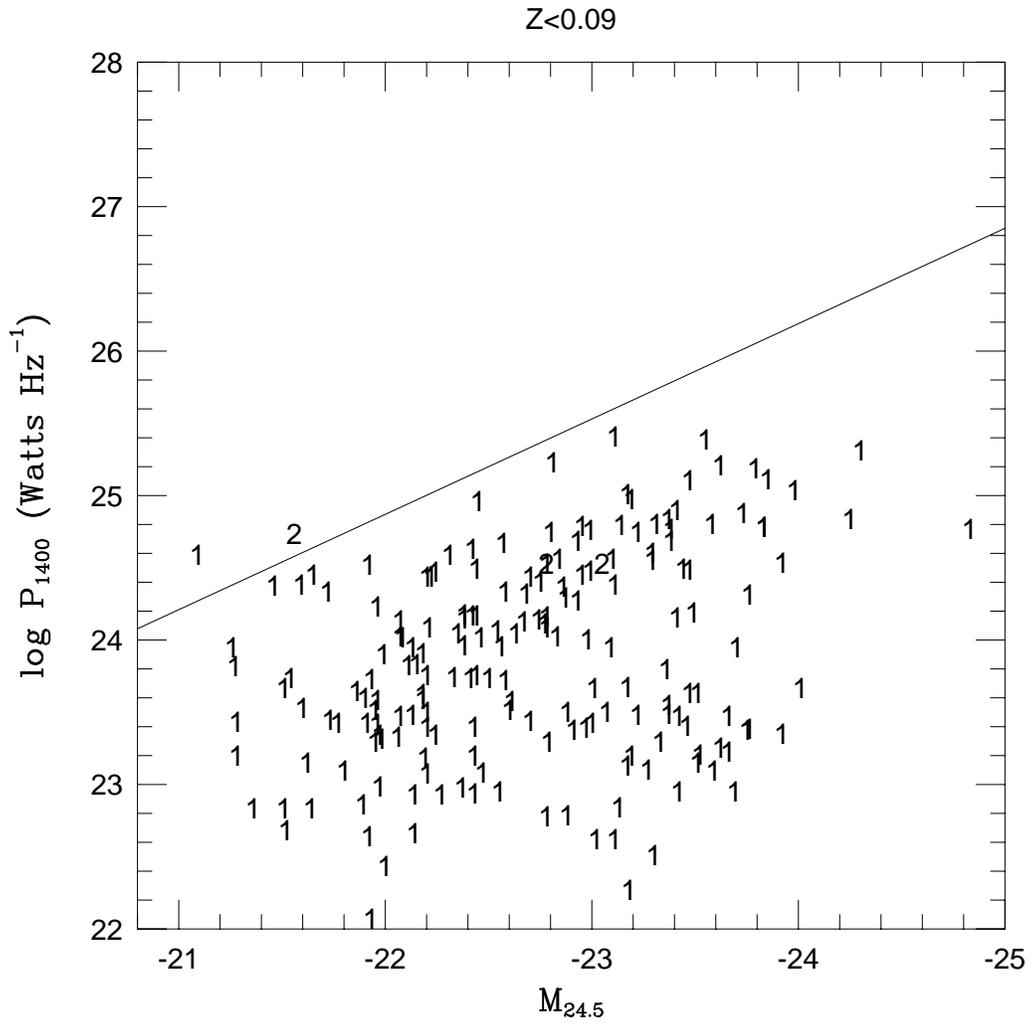}
\caption{FR I/II diagram for a statistically complete sample for $z<0.09$.}
\end{figure}

\clearpage

\begin{figure}[p]
\plotone{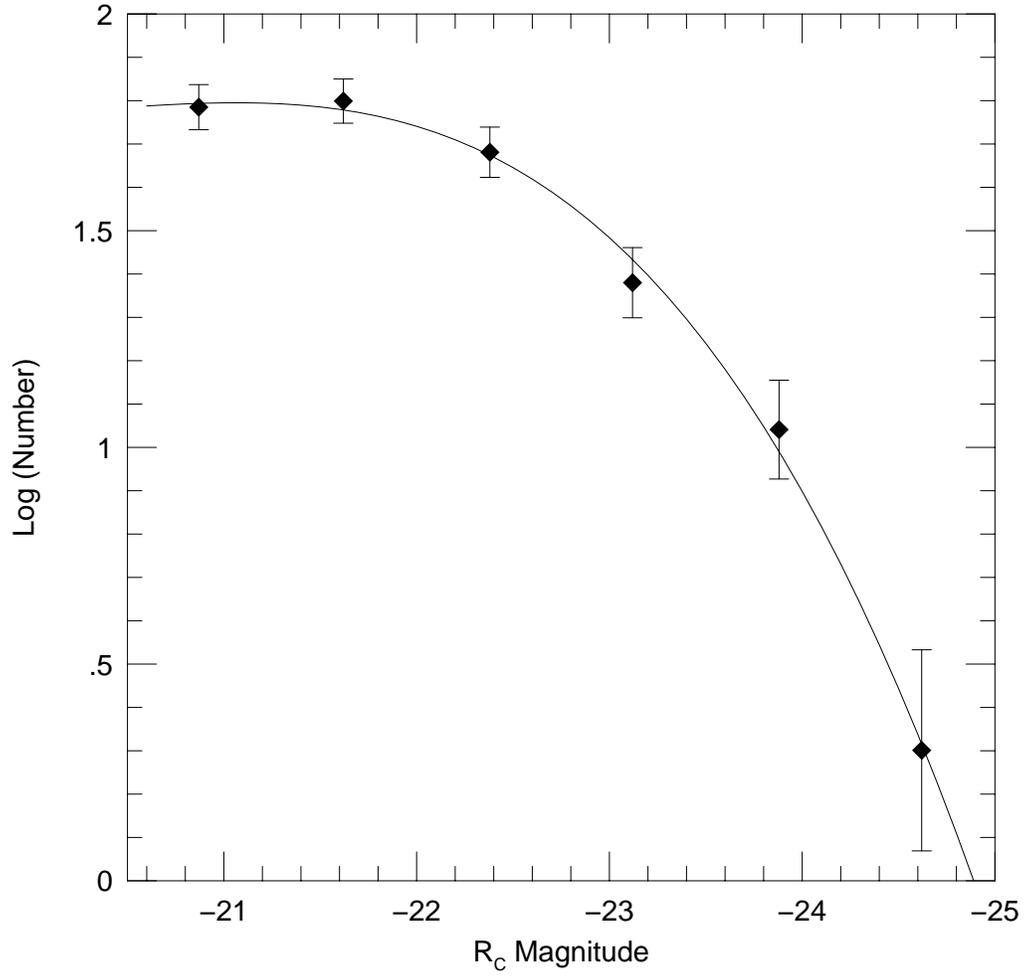}
\caption{Elliptical galaxy optical luminosity distribution for 25 clusters
from Dressler (1980).  The magnitudes have been converted to Cousins R ($R_C$).}
\end{figure}

\clearpage

\begin{figure}[p]
\plotone{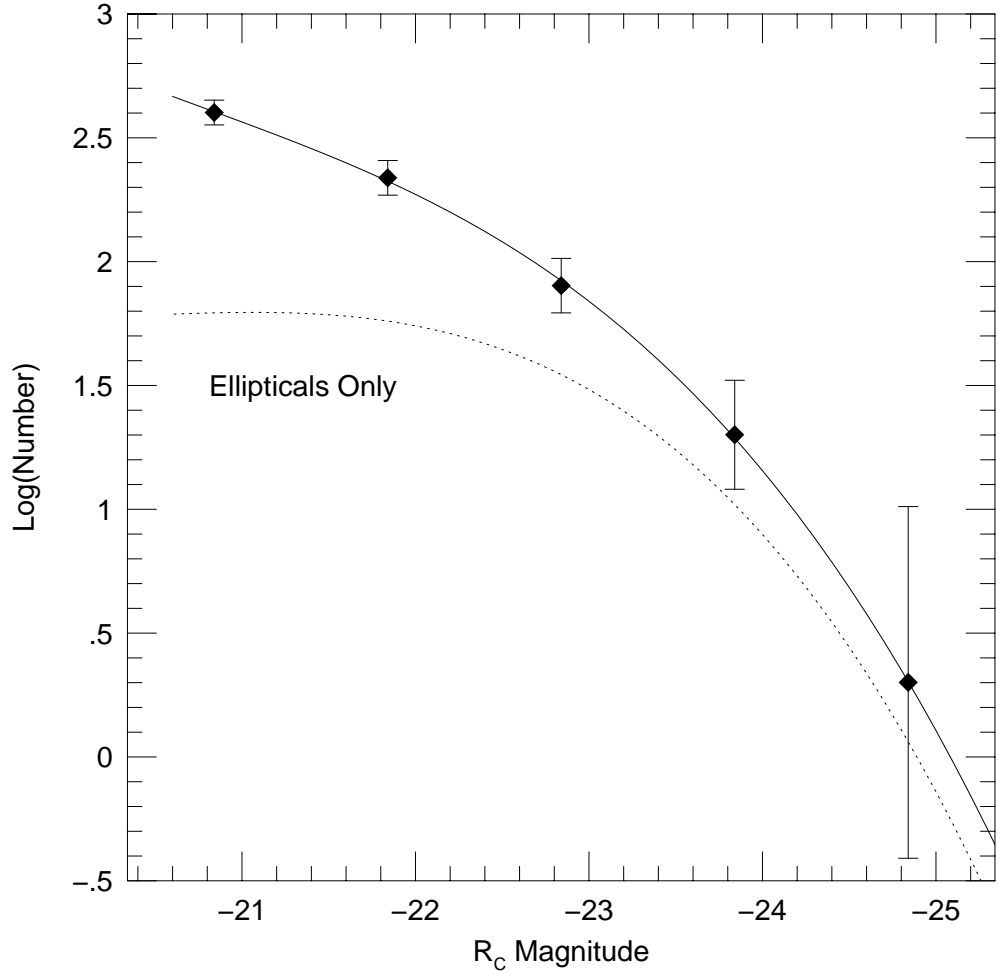}
\caption{Elliptical and S0 galaxy optical luminosity distributions, also
from Dresser (1980). The dotted line is reproduced from Figure 4 for elliptical 
galaxies only.}
\end{figure}

\clearpage

\begin{figure}[p]
\plotone{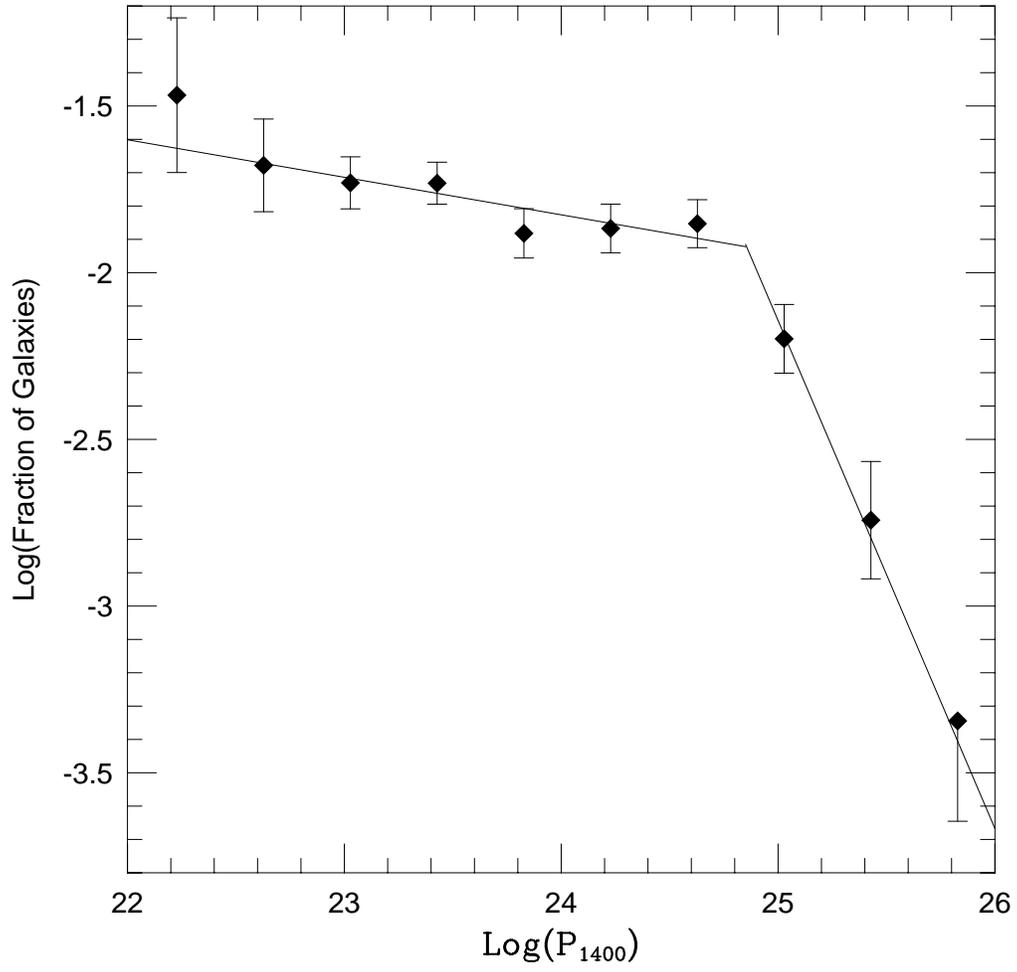}
\caption{The univariate radio luminosity function computed for our 
rich cluster sample with $z<0.09$.}
\end{figure}

\clearpage

\begin{figure}[p]
\plotone{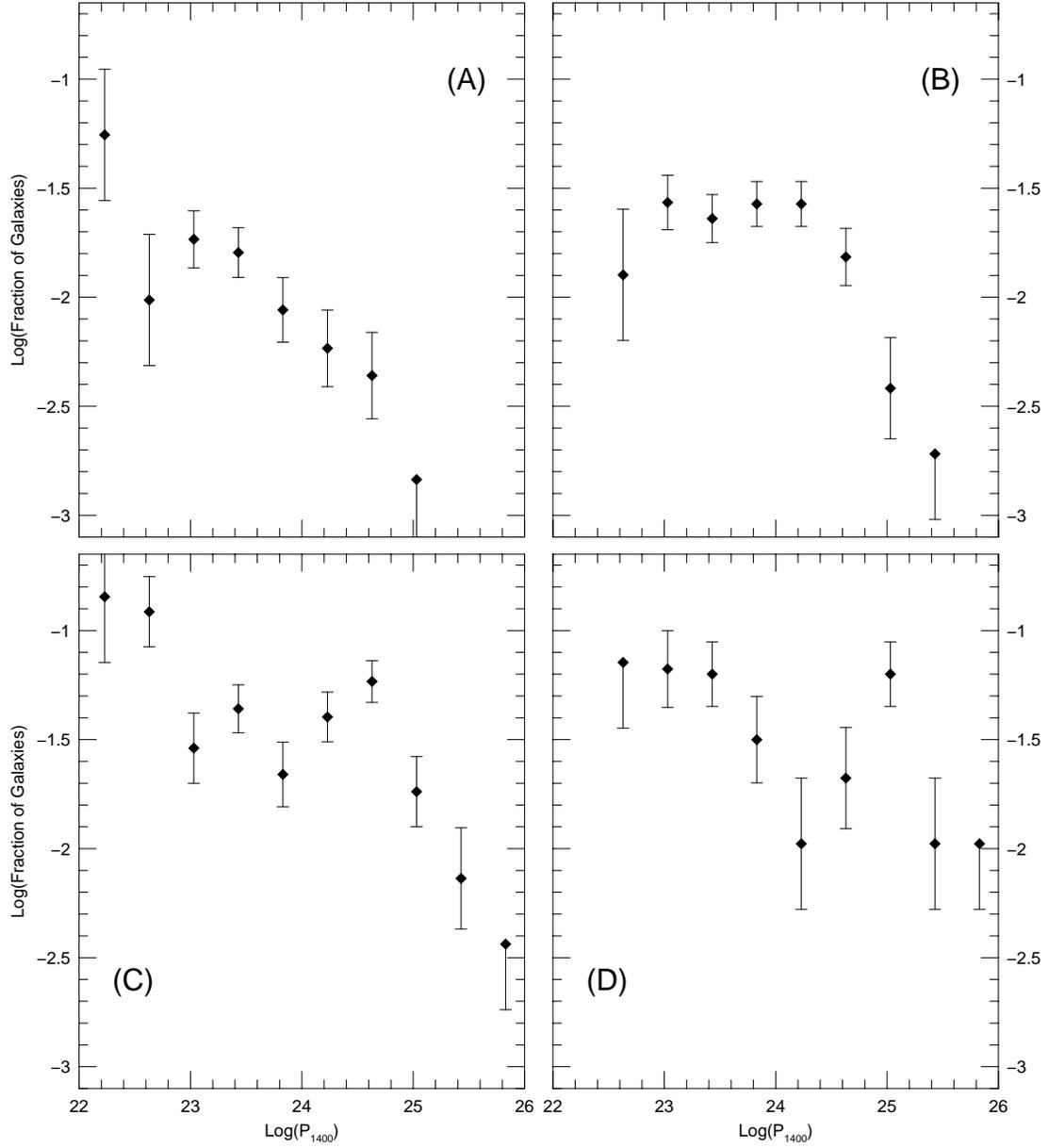}
\caption{Bivariate radio/optical luminosity functions for our statistical
sample.  The data is binned in $0.4\log P_{1400}$ and 0.75 magnitude 
intervals.   The bins are centered at (a) $M_{24.5}=-21.62$,  
(b) $M_{24.5}=-22.38$, (c) $M_{24.5}=-23.12$, and (d) $M_{24.5}=-23.88$.}
\end{figure}

\clearpage

\begin{figure}[p]
\plotone{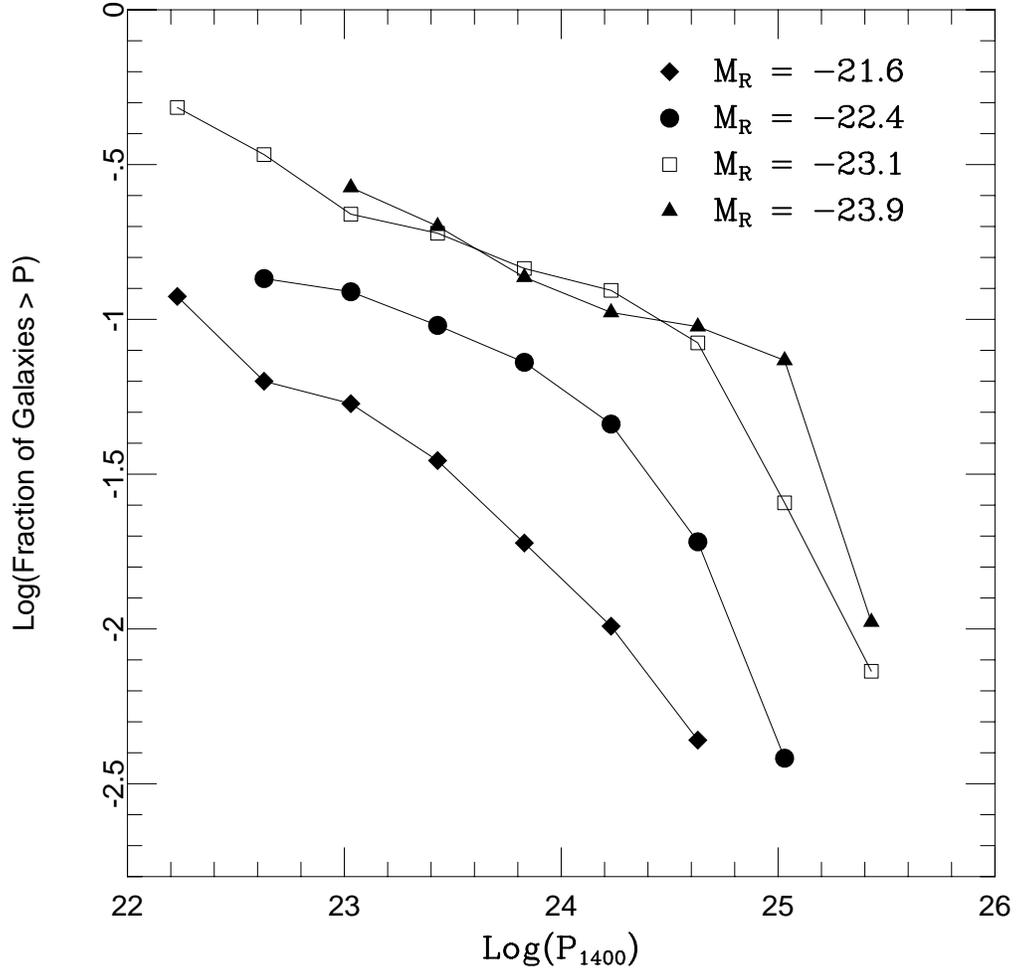}
\caption{The integral bivariate luminosity functions.  The magnitude
bins are labeled on the plot.} 
\end{figure}

\clearpage

\begin{figure}[p]
\plotone{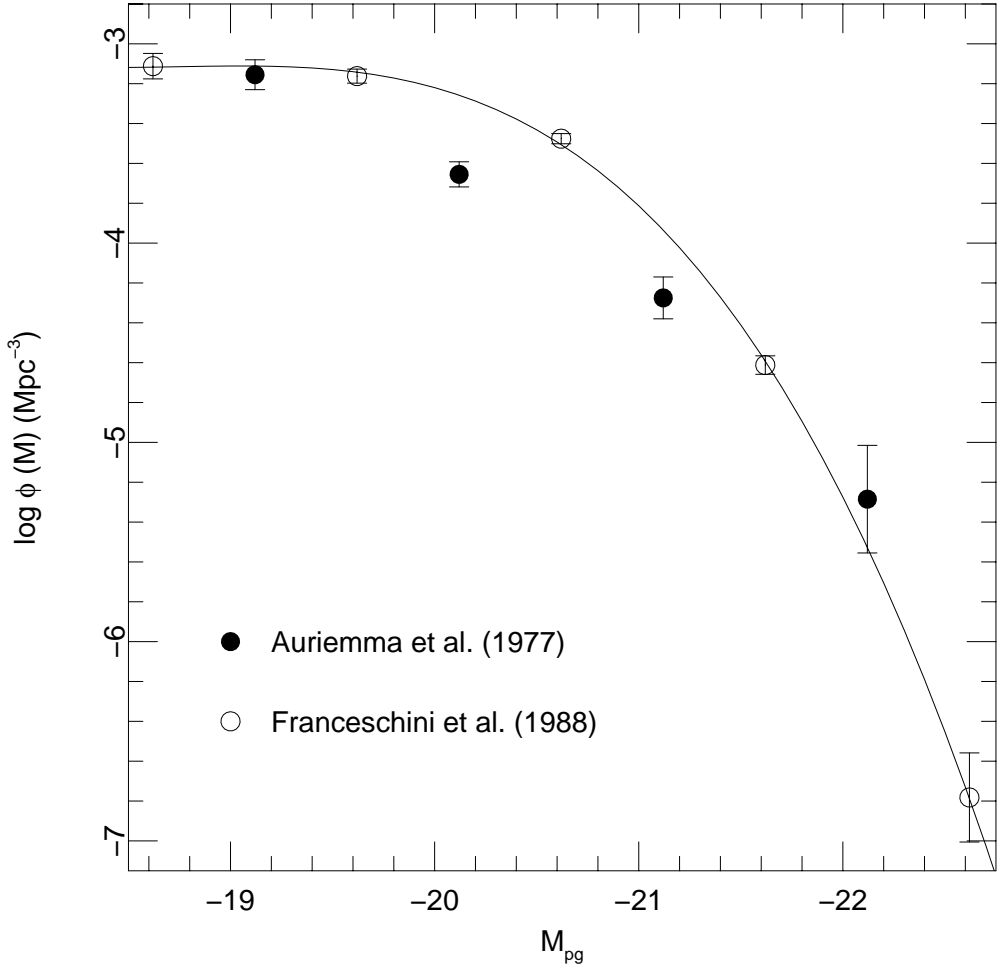}
\caption{The optical luminosity functions for E+S0 galaxies from 
Franceschini \etal\ (1988), and the function used by Auriemma \etal\ (1977).}
\end{figure}

\clearpage

\begin{figure}[p]
\plotone{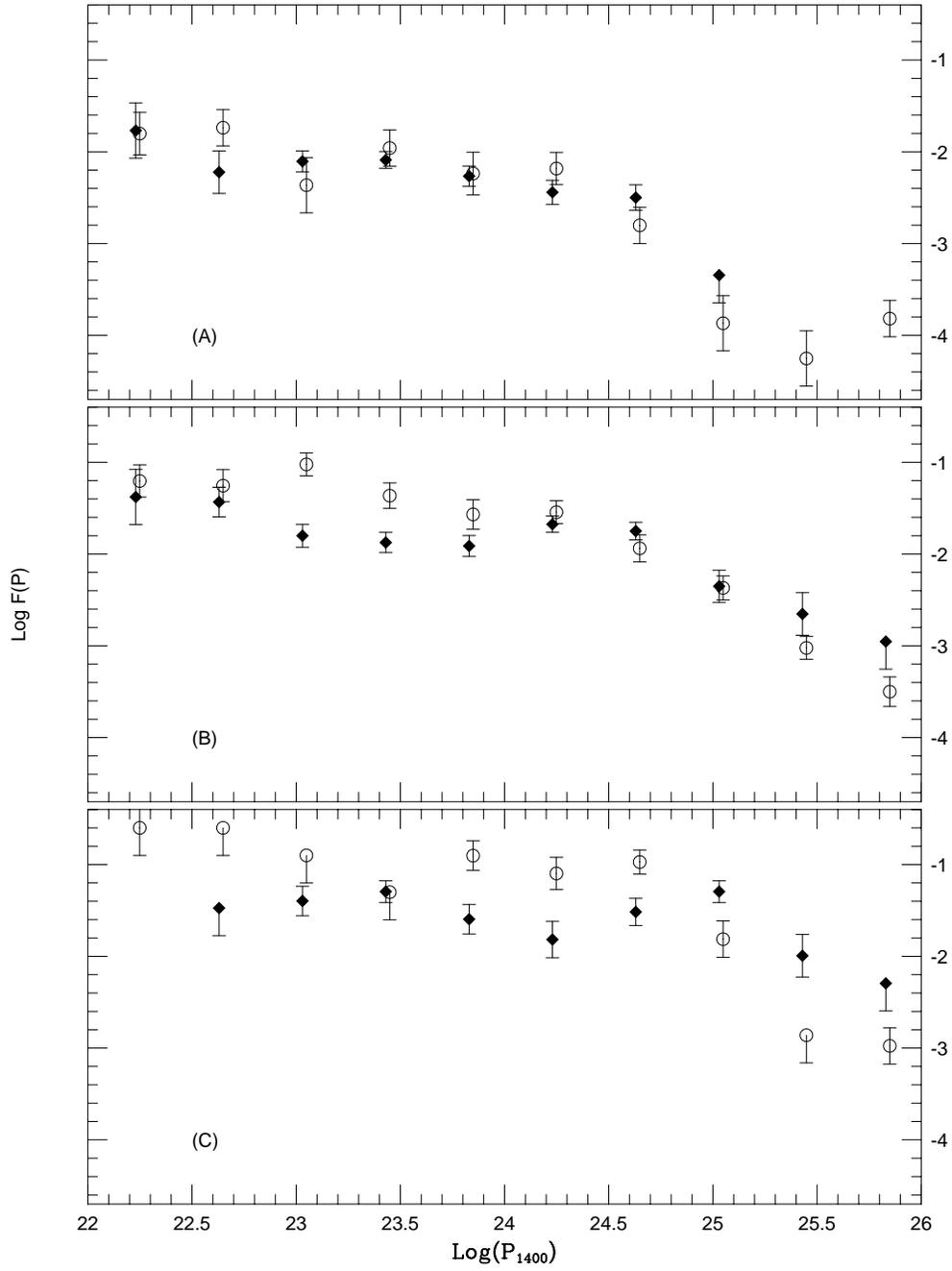}
\caption{The bivariate luminosity functions of Auriemma \etal\ 
(open circles) and 
our statistical cluster sample (solid diamonds).  The magnitude bins are 
(a) $M=-21.86$, (b) $M=-22.86$, and (c) $M=-23.86$.}
\end{figure}

\clearpage

\begin{figure}[p]
\plotone{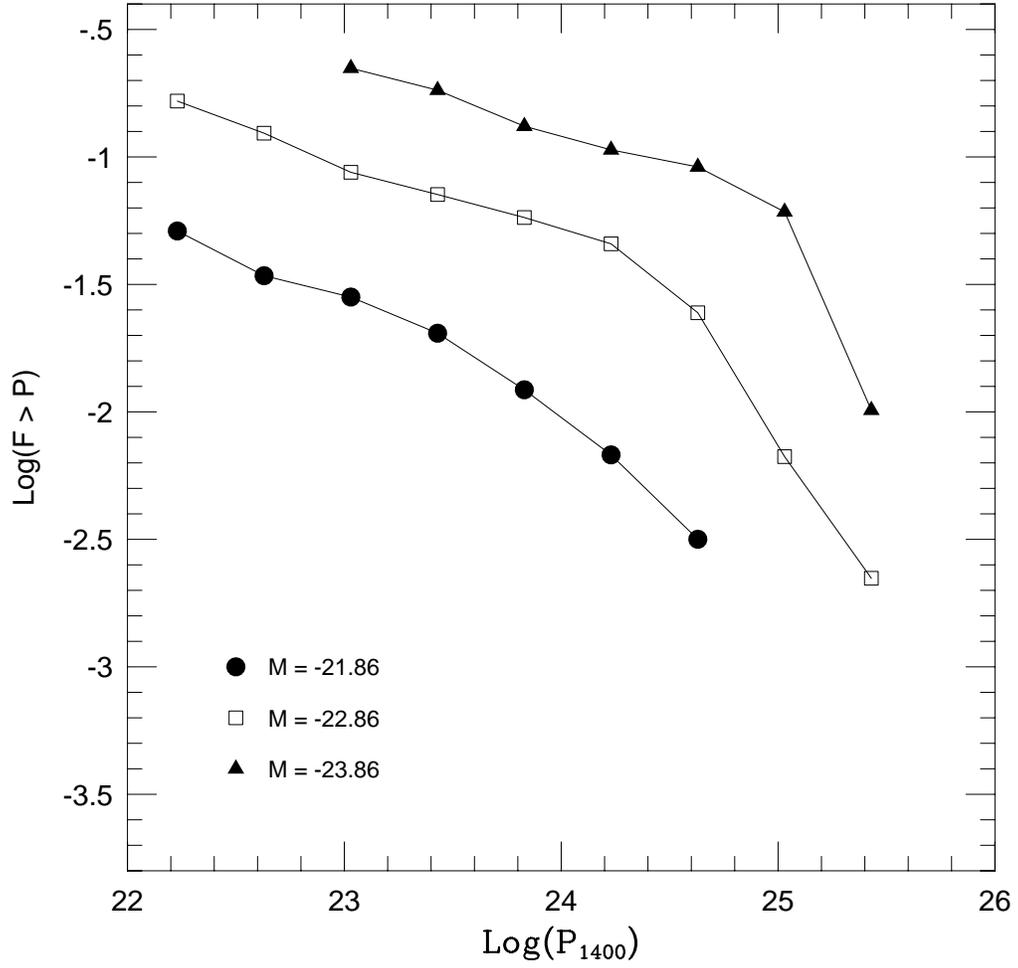}
\caption{The integral bivariate luminosity functions for our cluster 
sample binned in one magnitude intervals as in Auriemma \etal\ (1977).}
\end{figure}

\begin{figure}[p]
\plotone{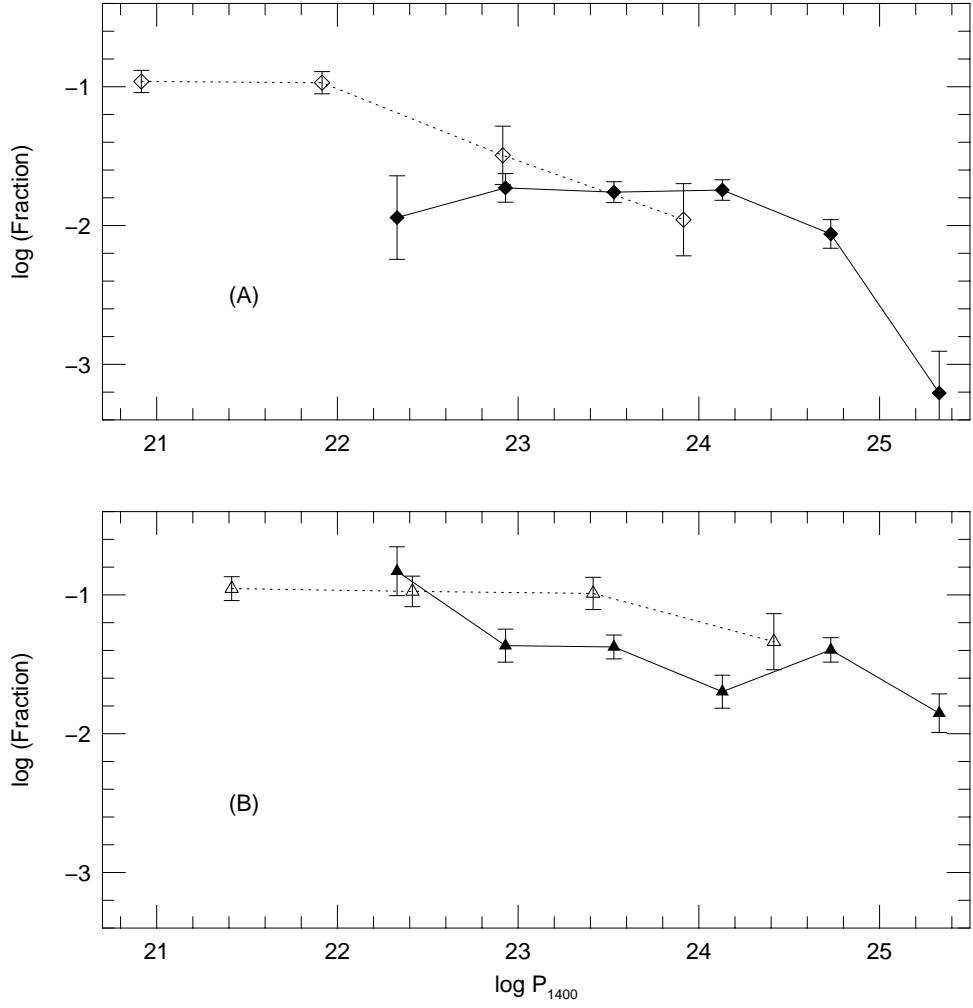}
\caption{The bivariate luminosity functions from our cluster sample 
(solid diamonds) and the sample from 
Sadler \etal\ (1989) (open diamonds).  The magnitude bins are 
(a) $M=-22.36$ and 
(b) $M=-23.36$.} 
\end{figure}

\end {document}